\documentclass[epj]{svjour}
\usepackage{graphics}
\usepackage{graphicx}
\begin{document}
\title{From quark drops to quark stars}
\subtitle{Some aspects of the role of quark matter in compact stars }
\author{Germ\'an Lugones 
}                     
%
%
\institute{Centro de Ci\^encias Naturais e Humanas, Universidade Federal do ABC, \\
            Rua Santa Ad\'elia, 166, 09210-170, Santo Andr\'e, Brazil}
\date{Received: date / Revised version: date}
%
\abstract{
We review some recent results about the mechanism of deconfinement of hadronic matter into quark matter in cold neutron stars and protoneutron stars.  We discuss the role of finite size effects and the relevance of temperature and density fluctuations on the nucleation process. We also examine the importance of surface effects for mixed phases in hybrid stars. 
A small drop of quark matter nucleated at the core of a compact star may grow if the conversion is sufficiently exothermic. In such a case, it may trigger the burning of the stellar core and even the whole star if quark matter is absolutely stable. We explore the physical processes that  occur inside the flame and analyze the hydrodynamic evolution of the combustion front.   
In the last part of this review, we focus on hybrid stars using the Nambu-Jona-Lasinio (NJL) model with scalar, vector and 't Hooft interactions,  paying particular attention to a generalized non-standard procedure for the choice of  the 'bag constant'.  
We also describe the non-radial oscillation modes of hadronic, hybrid and strange stars with maximum masses above  $2 M_{\odot}$ and show that  the frequency of the $p_1$ and $g$ fluid modes contains key information about the internal composition of compact objects. 
\PACS{
      {97.60.Jd}{neutron stars} \and 
      {25.75.Nq}{quark deconfinement}   \and
        {26.60.-c}{nuclear matter aspects of neutron stars} \and 
        {95.85.Sz}{gravitational waves} 
     } 
} 
\maketitle

\section{Introduction}

Observational evidence is not conclusive about the internal composition of compact stars. All measured stellar properties such as masses, radii, cooling properties,  etc. are reconcilable at present with a pure hadronic composition, but also with stellar models that allow the presence of deconfined quark matter. 
However, the recent discovery of two very high mass pulsars with $\sim 2 M_{\odot}$ \cite{Demorest2010,Antoniadis2013}
and  the possible existence of even more massive neutron stars  \cite{Romani2012,Kaplan2013} strengthens the idea that some kind of exotic matter must be present in such objects. 
In particular, since  the central baryon density in these heavyweight stars should be several times the nuclear saturation density, deconfined quark matter is a natural candidate for, at least, the core composition.

According to present knowledge,  quark matter could be formed if the density inside a purely hadronic star reaches some critical density. This may happen due to accretion onto a cold hadronic star in a binary system, as a consequence of cooling, deleptonization and fallback accretion during the protoneutron star phase of a just born compact star, due to spin down of a fast rotating star, or by binary merging. Other more exotic mechanisms such as strangelet contamination are also conceivable. The conversion of the star presumably begins with the nucleation of a small quark matter seed  which subsequently grows at the expenses of the gravitational energy extracted from the contraction of the object and/or through a strongly exothermic combustion process. All these scenarios lead to the formation of hot and neutrino rich quark matter occupying the core of a hybrid star or the whole compact object if quark matter is absolutely stable. Due to violent dynamics of the conversion, such event is expected to produce a conspicuous signal in neutrinos, gravitational waves and gamma rays.

In this work we review some recent results about the process of deconfinement of hadronic matter into quark matter in cold neutron stars and protoneutron stars. In Section 2, we  focus on the role of the surface tension and the curvature energy on the nucleation rate of quark drops, we present transition curves in the phase diagram and we discuss the relevance of temperature and density fluctuations on the nucleation process. We also analyze the importance of surface effects  for mixed phases in hybrid stars. 
An important feature of the nucleation process is that the drop must pass through an intermediate activation state before reaching chemical equilibrium.  Accordingly, the nucleation can be regarded as a two step process.  First, hadronic matter deconfines in a strong interaction timescale ($\sim 10^{-24}$ s) forming a quark matter drop out of chemical equilibrium with respect to weak interactions.  In spite of being very short lived, this is an unavoidable intermediate state that must be reached before arriving to the final configuration in chemical equilibrium. In the second step, weak interaction processes such as $u+d  \leftrightarrow  u+s$,  $u+e^{-}  \rightarrow  d+\nu_{e}$, $u+e^{-}  \rightarrow  s+\nu_{e}$,  drive the quark drop into an equilibrium state in a timescale of $\sim 10^{-8}$ s.  
The energy released in such conversion can trigger the conversion in the neighborhood of the drop  leading to the creation of a combustion front  that travels outwards along the star. In Section 3 we focus on the physical processes that  occur inside the burning flame and estimate the temperature rise and the neutrino emissivity. Then, we address the hydrodynamic evolution of the combustion front.  

In Section 4, we review a recent systematic study of hybrid star configurations using a relativistic mean-field hadronic equation of state and the Nambu-Jona-Lasinio (NJL) model with scalar, vector and 't Hooft interactions for three-flavor quark matter. We pay particular attention to the role of vector interactions and  to a generalized non-standard procedure for the choice of  the 'bag constant' that has a significant effect on the maximum mass of the hybrid configurations. 

In Section 5, we describe the non-radial oscillation spectrum of hadronic, hybrid and pure self-bound strange quark stars with maximum masses above $2 M_{\odot}$ using several equations of state. It is shown that the three kinds of objects present qualitatively different pulsating properties that allow to constrain the internal composition based on the frequency of the $f$, $p_1$ and $g$ modes. We conclude with a brief summary in Section 6.

\section{Finite size effects in quark matter}  

Finite size effects are extremely relevant for neutron star physics. Not only  they  determine whether  a mixed phase of quarks and hadrons may appear inside hybrid stars, but also have a key role in the beginning of the conversion of a hadronic star into a quark star, since they regulate the nucleation rate of the initial quark drops that trigger the conversion.  In this section we review some recent results obtained within the multiple reflection expansion (MRE)  formalism  developed by Balian and Bloch \cite{Balian1970}.

\subsection{Summary of the multiple reflection expansion formalism}

In a confined droplet of massive quarks,  surface tension arises dynamically due to the modification of the density of quark states due to the boundary conditions.  The density of states $\rho \equiv dN/dk$ of a degenerate Fermi gas  in a spherical volume $V =4/3 \pi R^3$ with surface area $S=4 \pi R^2$ and curvature $C = 8 \pi R$ can be written as \cite{Balian1970}:
\begin{equation}
\rho_{MRE}(k,m_f,R) = 1 + \frac{6\pi^2}{kR} f_S + \frac{12\pi^2}{(kR)^2} f_C
 \end{equation}
being $f_S  = - \frac{1}{8 \pi} [1 -\frac{2}{\pi} \arctan (k/m_f) ]$ the surface contribution  and 
$f_C  =  \frac{1}{12 \pi^2} [1 -\frac{3k}{2m_f} \left(\frac{\pi}{2} - \arctan (k / m_f) \right)]$ the curvature one \cite{Madsen1994}, where $m_f$   is the mass of the different quark flavors ($f=u, d, s$).  Since  $m_s$ is significantly larger than $m_u$ and $m_d$ the main surface corrections come in general from the contribution of strange quarks. 

To introduce finite size effects in a given equation of state, the integral that defines the thermodynamic potential of the model, is modified according to the generic replacement  \cite{Madsen1994,Kiriyama2003,Kiriyama2005,Lugones2011,Lugones2013}
\begin{equation}
\int  {{\cdots}} \frac{k^2 \, dk}{2 \pi^2}  \longrightarrow
\int {{\cdots}} \frac{k^2 \, dk}{2 \pi^2} \rho_{MRE}.
\label{MRE}
\end{equation}
For massive quarks, the density of states  is reduced compared with the bulk one, and for a 
range of small momentum becomes negative. Such non-physical  values are removed
by introducing an infrared cutoff  $\Lambda_{IR}$ in momentum space, where 
 $\Lambda_{IR}$ is the largest solution of the equation $\rho_{MRE} (k)= 0$ with respect to the momentum $k$. 

The  thermodynamic potential obtained after the above replacement can always be written as the sum of a volume term, a surface term and a curvature term:
\begin{equation}
\Omega_{{MRE}} = -P V + \alpha S + \gamma C ,
\label{eq17}
\end{equation}
where  $P$ is the pressure, $\alpha$ is the surface tension and $\gamma$ is  the curvature energy density (see e.g.  \cite{Lugones2011} and references therein).

The explicit form of $P$, $\alpha$ and $\gamma$ depends on the specific EOS. For example, we can use  an $SU(3)_f$ NJL effective model  including color superconducting quark-quark interactions, whose Lagrangian is given by 
\begin{eqnarray}
{\cal L} &=& \bar \psi \left(i \rlap/\partial - \hat m \right) \psi \nonumber \\
& + &  G \sum_{a=0}^8 \left[ \left( \bar \psi \ \tau_a \ \psi \right)^2 + \left( \bar \psi \ i \gamma_5 \tau_a \ \psi \right)^2 \right]
\label{lagrangian} \\
& + &  2H \!\! \sum_{A,A'=2,5,7} \left[ \left( \bar \psi \ i \gamma_5 \tau_A \lambda_{A'} \ \psi_C \right) \left( \bar \psi_C \ i \gamma_5 \tau_A \lambda_{A'} \ \psi \right) \right] \nonumber
\end{eqnarray}
where $\hat m=\mathrm{diag}(m_u,m_d,m_s)$ is the current mass matrix in flavor space, the matrices $\tau_i$ and $\lambda_i$ with $i=1,..,8$ are the Gell-Mann matrices corresponding to the flavor and color groups respectively, and $\tau_0 = \sqrt{2/3}\ 1_f$. 

In this case, the expressions for $\alpha$ and $\gamma$ are \cite{Lugones2011,Lugones2013}:
\begin{equation}
\alpha \equiv  \frac{\partial \Omega_{{MRE}}}{ \partial S }
\bigg|_{T, \mu, V, C}  = 2 \int_{{{\Lambda_{IR}}}}^\Lambda k \;
dk \; f_S \sum_{i=1}^9 \omega(x_i,y_i) ,
\label{surfacetension}
\end{equation}
\begin{equation}
\gamma \equiv  \frac{\partial \Omega_{{MRE}}}{ \partial C } \bigg|_{T, \mu, V, S}  = 2 \int_{{{\Lambda_{IR}}}}^\Lambda  dk \; f_C \sum_{i=1}^9 \omega(x_i,y_i).
\label{curvatureenergy}
\end{equation}
where $\Lambda$ is the cut-off of the model and $\omega(x,y)$ is defined by $\omega(x,y) =  - x - T \ln[1+e^{-(x-y)/T}]  -  T \ln[1+e^{-(x+y)/T}] $. The quantities $x$ and $y$ depend on the momentum $k$, the chemical potentials $\mu_{cf}$, the pairing gaps $\Delta_i$ and the chiral condensates $\sigma_i$ (for more details see  \cite{Lugones2011,Lugones2013}).

As shown above, $\alpha$ and $\gamma$ depend on several variables such as  $\mu_{cf}$,  $\Delta_i$ and $\sigma_i$,  that are not independent but must obey several constrains. For the NJL model shown above, the consistent solutions of the model correspond to the stationary points of  the thermodynamic potential $\Omega$ with respect to  $\sigma_i$ and $\Delta_i$; i.e., we have ${\partial\Omega}/{\partial\sigma_i} =0$ and  ${\partial\Omega} / {\partial|\Delta_i|} = 0$.
On the other hand, the chemical potentials fulfill conditions that depend on  the physical situation under study. For example, in the case of quark droplets in a hypothetical mixed phase present in a hybrid star the quark species are in equilibrium under weak interactions and the $\mu_{cf}$ are subject to a set of chemical equilibrium conditions. In contrast, during the nucleation of the first quark matter drop that triggers the conversion to quark matter in a pure hadronic star, weak interactions are frozen. In such a case, the $\mu_{cf}$ are related by flavor conservation conditions.

\subsection{Finite size effects and the formation of a quark star: nucleation of quark drops}

We focus here on the problem of the nucleation of a small quark matter drop near the center of a hadronic star. We want to determine the conditions under which such nucleation is energetically favored, e.g. at which density and temperature it may occur. Later we will focus on the problem of weather the quark drop can grow and convert a significant part of the hadronic star into quark matter.

\begin{figure}[t]
\begin{center}
\resizebox{0.5 \textwidth}{!}{ \includegraphics[angle=0]{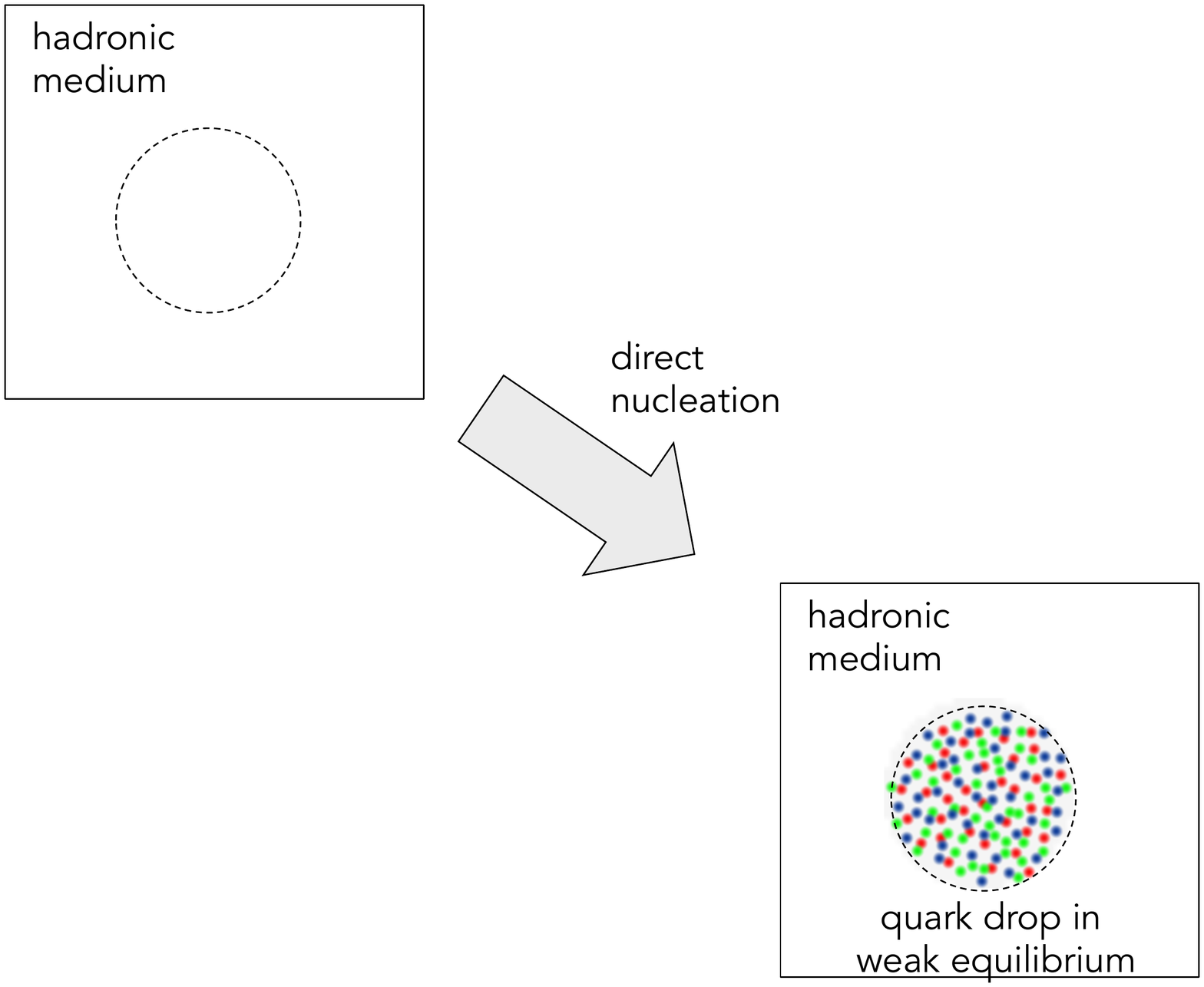}    }
\caption{The direct nucleation of a  quark matter drop in equilibrium under weak interactions is a high order weak process which is strongly suppressed by a factor $\sim G_F^{2N/3}$, because, for a drop containing $N$ baryons, it involves the simultaneous conversion of roughly $N/3$ non-strange quarks into strange quarks.}
\label{fig_deconf_1}
\end{center}
\end{figure}

\begin{figure}[t]
\begin{center}
\resizebox{0.5 \textwidth}{!}{ \includegraphics[angle=0]{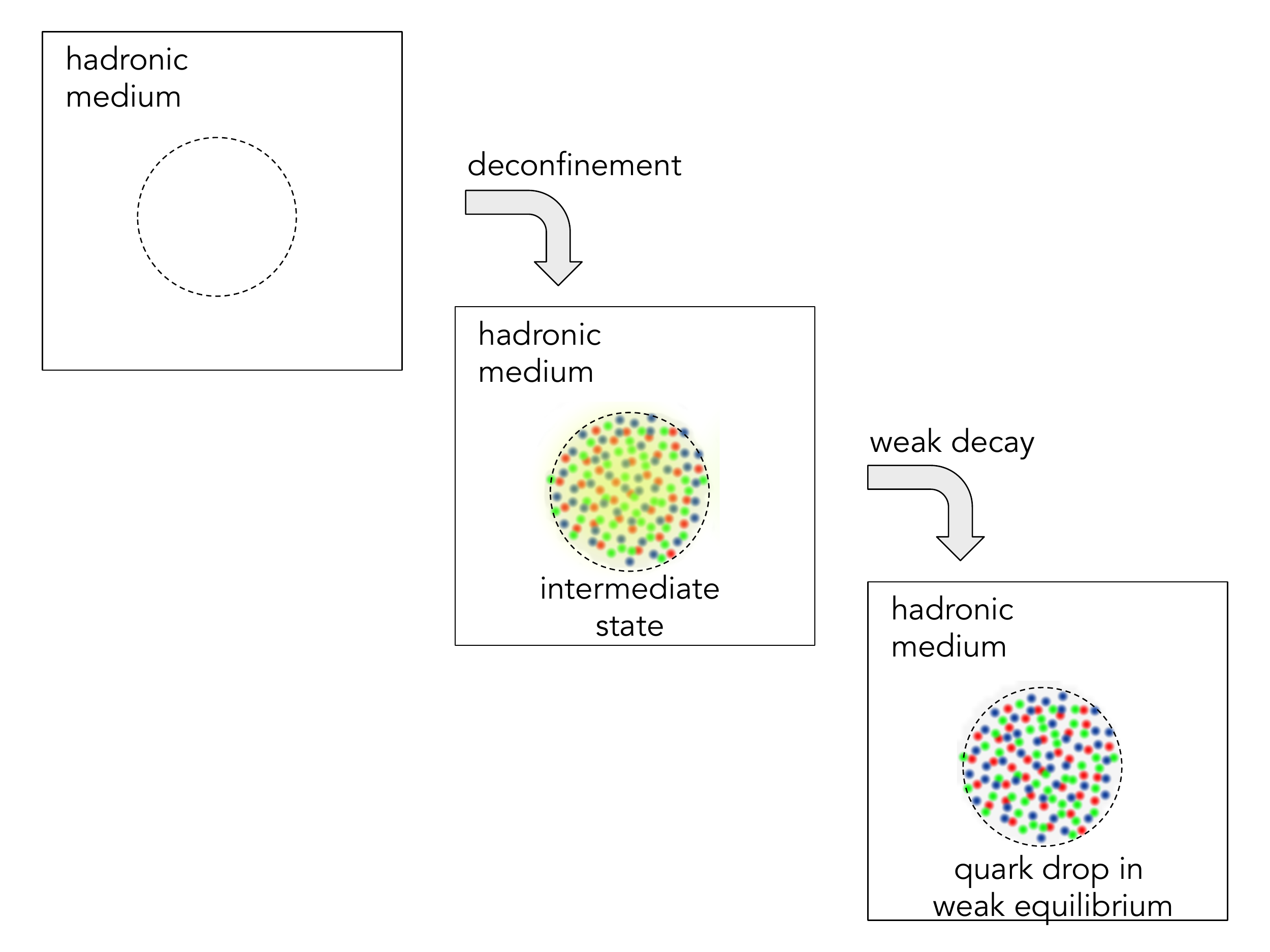}    }
\caption{The nucleation can proceed through and intermediate state.  In the first step, hadronic matter deconfines in a strong interaction timescale ($\sim 10^{-24}$ s) and a quark matter drop out of chemical equilibrium with respect to weak interactions is formed.  Such drop is very short lived, but it constitutes an unavoidable intermediate state that must be reached before arriving to the final configuration in chemical equilibrium. In the second step, weak interaction processes such as $u+d  \leftrightarrow  u+s$,  $u+e^{-}  \rightarrow  d+\nu_{e}$, $u+e^{-}  \rightarrow  s+\nu_{e}$,  etc., drive the quark drop into an equilibrium state in a timescale of $\sim 10^{-8}$ s.  }
\label{fig_deconf_2}
\end{center}
\end{figure}

An important aspect of the nucleation process is that the direct formation of a quark matter droplet in chemical equilibrium under weak interactions is strongly suppressed (see Fig. \ref{fig_deconf_1}).  Instead, the nucleation can start through an intermediate state out of chemical equilibrium (see Fig. \ref{fig_deconf_2}). The initial flavor composition of the intermediate state is determined by the conservation of quark and lepton flavors with respect to the hadronic environment (see \cite{Lugones2011,Lugones2010,Lugones2009,doCarmo2013b} and references therein).  Clearly, in a timescale of $\sim 10^{-8} \textrm{s}$  the quark droplet will reach equilibrium with respect to weak interactions.

To model the intermediate state we may assume that the drop is spherical and that  it is in thermal and  mechanical equilibrium with the hadronic medium, i.e.,  $T^H = T^Q$ and  $P^Q  - \frac{2 \alpha}{R}   -   \frac{2 \gamma}{R^2}  - P^H  = 0$. Additionally,  the Gibbs free energy per baryon must be  the same for both hadronic  and quark matter, $g^H = g^Q$.   
Also, since weak interactions are frozen,  flavor is conserved  and the particle abundances must be the same in  hadronic matter and in the intermediate deconfined quark drop. This can be written as
\begin{equation}
Y^H_f = Y^Q_f   \;\;\;\;\;\; f=u,d,s,e, \nu_e \label{flavor}
\end{equation}
where $Y^H_f \equiv n^H_f / n^H_B$ and  $Y^Q_i \equiv n^Q_f / n^Q_B$ are the abundances of each particle in the hadron and quark phase respectively, being $n_f$ the particle number density and $n_B$ the baryon number density.  Notice that, since the hadronic phase is assumed to be electrically neutral, flavor conservation ensures automatically the charge neutrality of the just nucleated quark drop.  
Finally, it has been shown in \cite{Lugones2005} that when color superconductivity is included in the analysis together with flavor conservation,  the most likely configuration of the just deconfined phase is a two-flavor color superconductor (2SC) provided the pairing gap is large enough.

\begin{figure}[b]
\begin{center}
\resizebox{0.5 \textwidth}{!}{ \includegraphics[angle=-90]{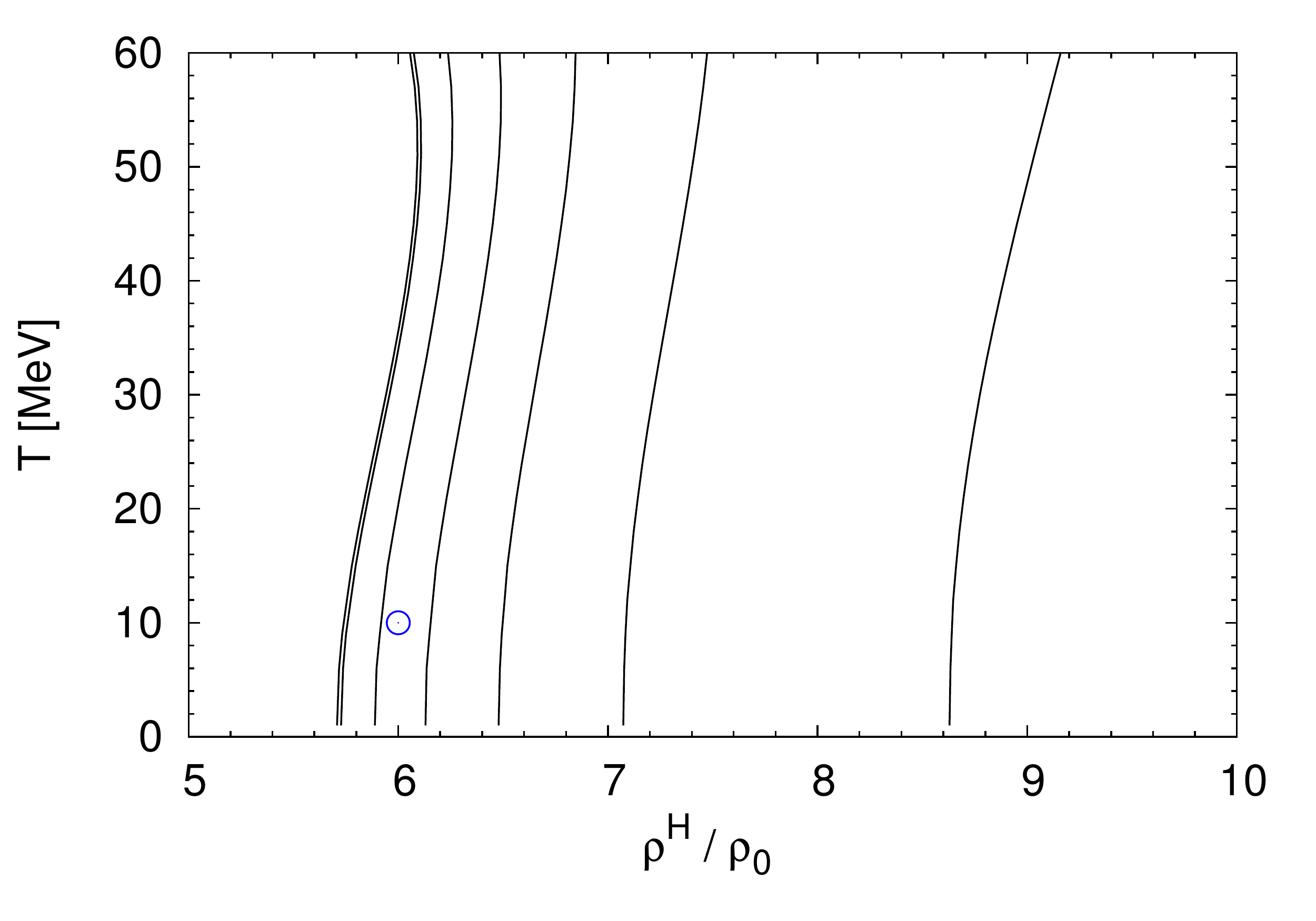}    }
\caption{Transition curves for different radii.   From right to left the curves correspond to $R [\textrm{fm}]= 2, 5, 10, 20, 50, 500, \infty$.  The curve for the bulk case ($R= \infty$) is almost coincident with the curve for $R = 500$ fm. For hadronic matter we used a relativistic mean-field  equation of state \cite{Walecka1974,Glendenning1991} and fixed $\mu_{\nu_e}^H = 200$ MeV. For quark matter we have employed the NJL model presented in Eq.  (\ref{lagrangian}).  For more details see \cite{Lugones2011}.  }
\label{transition_curves}
\end{center}
\end{figure}

Within the above model,  all the thermodynamic properties of the intermediate state can be obtained once we fix the radius $R$ of the deconfined drop for a given temperature $T$ and neutrino chemical potential of the trapped neutrinos in the hadronic phase $\mu_{\nu_e}^H$. In particular,   for each set $\{R, T, \mu_{\nu_e}^H \}$  there is a unique hadronic mass energy density $\rho^H$ at which the deconfinement conditions are fulfilled. This allows the construction of  ``transition curves'' in the  density versus temperature plane as show in Fig. \ref{transition_curves}.

For interpreting the transition curves, let us consider the following example:
suppose that a portion of hadronic matter inside a neutron star core  has a density $\rho^H
\approx 6 \times \rho_0$, a temperature $T = 10$  MeV,  with
trapped neutrinos having a chemical potential $\mu_{\nu_e}^H =
200$ MeV. This state is represented by a circle in Fig. \ref{transition_curves}. 
Comparing the position of the circle with
respect to the curves for different radii, we see
that it is energetically favorable to convert hadronic matter into
quark drops provided they have a radii larger than $R \sim 50$ fm. That is, radii whose curves lie to the left
of the circle are energetically favored while those on the right are forbidden because of the energy cost associated with surface and curvature.  Such energy cost is linked with the values of $\alpha$ and $\gamma$, which for the NJL model of Eq.  (\ref{lagrangian}) are in the range  $\alpha = 120 - 150$ MeV fm$^{-2}$ and  $\gamma = 100 -110$ MeV fm$^{-1}$  for 
$\{R, T, \mu_{\nu_e}^H \} = \{2 - \infty \, \textrm{fm}, 0-60 \, \textrm{MeV},  0-200 \, \textrm{MeV} \} $.

\begin{figure}
\begin{center}
\resizebox{0.45 \textwidth}{!}{\includegraphics[angle=-90]{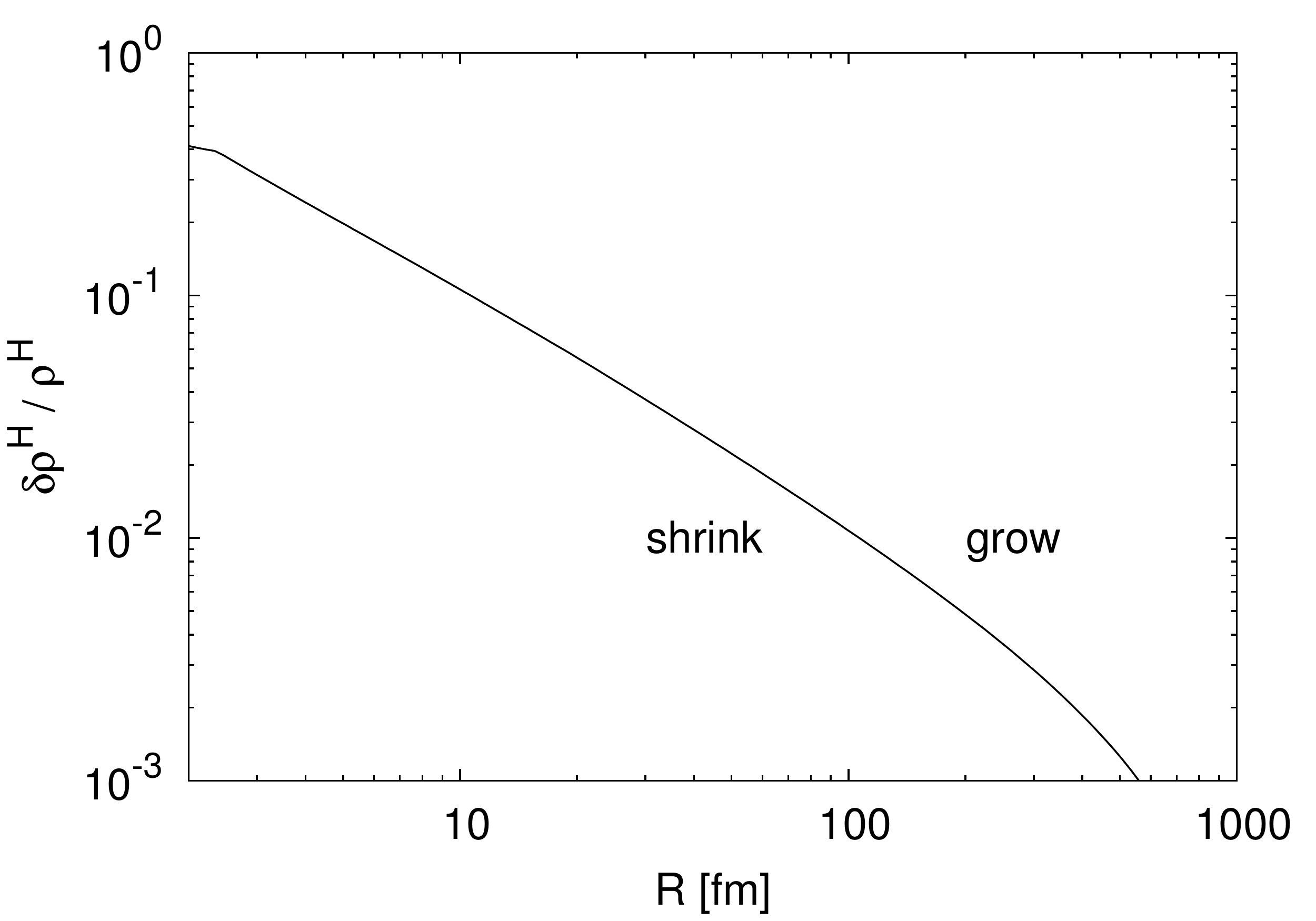} }
\caption{Critical curve for fluctuations in hadronic matter.  Mass-energy densty fluctuations in hadronic matter having a given $\delta \rho^H / \rho^H$ are able to grow if they have a size $R$ larger than the here-shown critical one.
The curve has been calculated for T= 2 MeV and $\mu_{\nu_e}^H = 0$, but the result does not depend significantly on the temperature and the chemical potential of trapped neutrinos \cite{Lugones2011}.   }
\label{spectrum}
\end{center}
\end{figure}

The transition curves shown in Fig. \ref{transition_curves} allow an analysis of the relevance of thermodynamic fluctuations at the core of a hadronic neutron star.   Let us consider, for simplicity, a region of hadronic matter at the stellar core  having a chemical potential $\mu_{\nu_e}^H$ and a mass-energy density  \textit{infinitesimally to the right} of the curve with $R = \infty$ in the $T$ versus $\rho^H/\rho_0$ plane corresponding to the same  $\mu_{\nu_e}^H$. Since this density corresponds to the bulk transition we shall refer to it as $\rho_{bulk}^H(T,\mu_{\nu_e}^H)$.   
For such a density, it is  favorable to convert the system into quark matter \textit{in bulk}, but this is not possible in practice due to the surface and curvature energy cost. 
However, fluctuations in the independent thermodynamic
variables $\{T, \rho^H,\mu_{\nu_e}^H \}$ of the hadronic fluid may
drive some portion of it  to a state described by $\{T + \delta T,
\rho^H + \delta \rho^H,\mu_{\nu_e}^H + \delta \mu_{\nu_e}^H  \}$.
Nevertheless, notice that the curves of Fig. \ref{transition_curves}  are quite vertical,
i.e. the transition is not very sensitive to changes in $T$. The
same holds for variations in $\mu_{\nu_e}^H$ as can be checked in \cite{Lugones2011}. Thus, we shall
consider only energy-density fluctuations with radius $R_{fl}$
that drive some part of the hadron fluid to a density
$\rho_{*}^{H}  = \rho_{bulk}^H + \delta \rho^H$ to the right of a
curve with a given $R$. As explained above, quark drops with radii larger than $R$ are energetically favored; i.e., if the fluctuation has a size $R_{fl}$ larger than $R$ it will be energetically favorable for it to grow indefinitely.  In
order to quantify this, we calculate from the transition curves the difference $\delta
\rho^H$ between $\rho_{bulk}^H$ and the hadronic density of the
point that allows nucleation with radius $R$ or larger. In such a
way we can construct  a critical spectrum  $\delta \rho^H/\rho^H$
as a function of $R$ for different values of $T$ and
$\mu_{\nu_e}^H$ as seen in Fig. \ref{spectrum}. Fluctuations of a given
over-density $\delta \rho^H/\rho^H$ must have a size larger than
the critical value given in Fig. \ref{spectrum} in order to grow. Equivalently,
fluctuations of a given size must have an over-density $\delta
\rho^H/\rho^H$ larger than the critical one for that size.

\begin{figure}
\begin{center}
\resizebox{0.45 \textwidth}{!}{
\includegraphics[angle=-90]{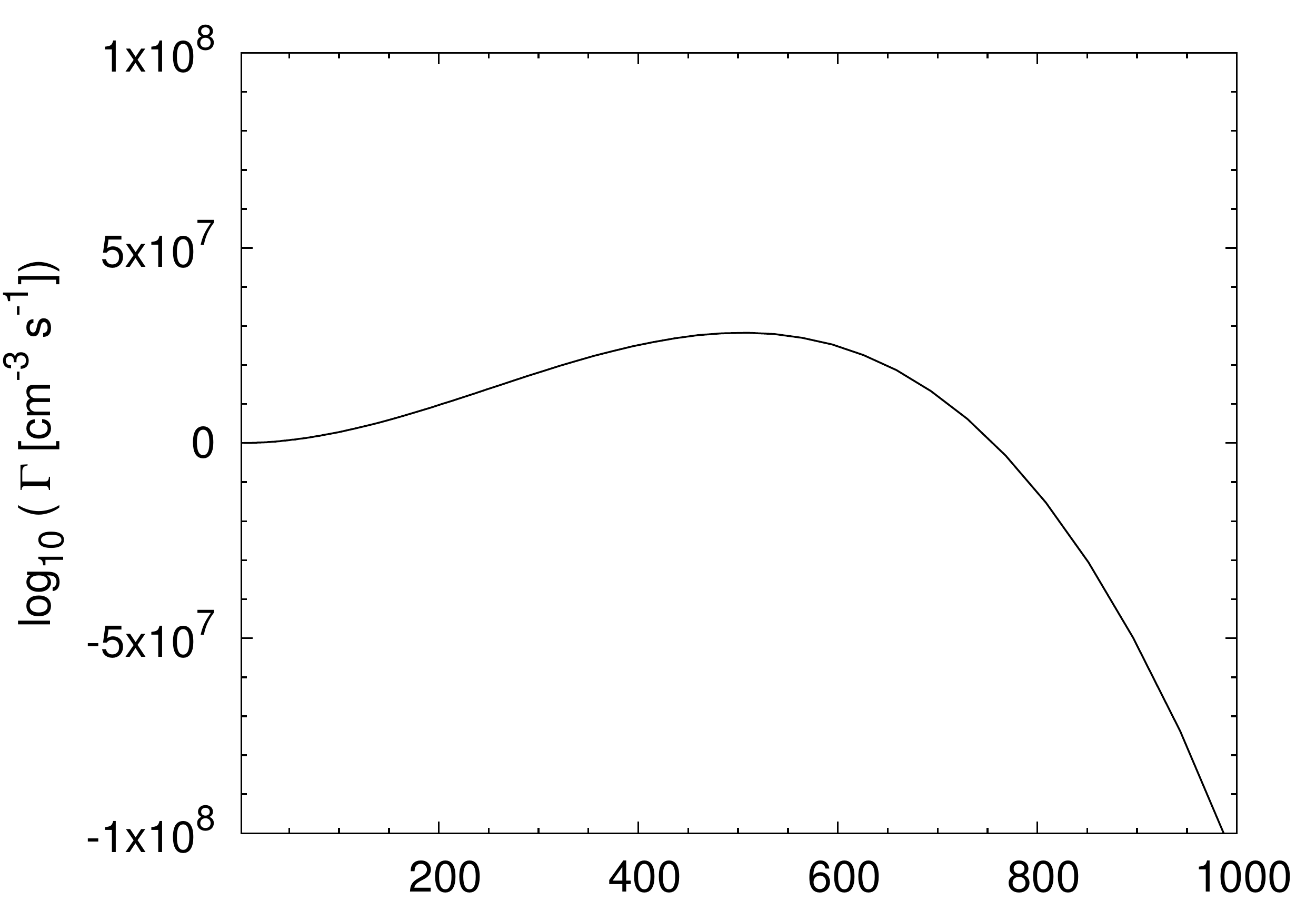}
}
\caption{Nucleation rate for bubbles of the critical size.   }
\label{rate}
\end{center}
\end{figure}

The nucleation rate of critical bubbles can be be determined through
\begin{equation}
\Gamma \approx T^4 \exp (-\delta \Omega_c/T).
\label{prob}
\end{equation}
where  $\delta \Omega_c$ is the work required to transform a portion of hadronic matter at the bulk transition point into a quark bubble with the critical radius 
\begin{equation}
\delta \Omega_c \equiv   - {{4 \pi} \over 3} R^3 (P^Q - P^H_{bulk} ) + 4 \pi \alpha R^2  + 8 \pi \gamma R .
\label{wcrit}
\end{equation}
The  $T^4$ prefactor is rather arbitrary and is included here on dimensional grounds. However,  $\Gamma$ is largely dominated by the exponent in Eq. (\ref{prob}); i.e. we always have $\log_{10} \Gamma \approx \log_{10} (\mathrm{prefactor})  -  \delta \Omega_c/[ T  \ln(10)]$ where the second term is much larger than the first one.  
A more elaborate statistical prefactor has been developed  in the literature \cite{prefactor}. Nevertheless, the determination of such prefactor involves the specification of transport coefficients such as the thermal conductivity, and the shear and bulk viscosities, which are not well known for ultradense matter (see Refs. \cite{Logoteta2009,prefactor} for more details). We have checked that the prefactor given in Ref. \cite{Logoteta2009} is in fact very different from the  $T^4$ factor, but it is not dominant with respect to the exponential for the conditions encountered in our calculations. Thus, the results are not significantly affected by that choice. 
The nucleation rate is given in Fig. \ref{rate} and shows that critical bubbles with $R >  800$ fm are strongly disfavored while those with $R <  800$ fm have a huge rate. The nucleation time $\tau$ is the typical time needed to nucleate a droplet of radius $R$ and is given by $\tau =( \frac{4}{3} \pi R^3 \ \Gamma )^{-1}$.

The above presented  results are in qualitative agreement with similar calculations  using the MIT bag model with color superconductivity for quark mattter \cite{doCarmo2013}. Quantitatively, the  nucleation rate obtained within the MIT bag model is very different to the one obtained within the NJL model; nonetheless, both are negligible above some radius $R^*$ and huge below it. Moreover,  in spite of the large differences in $\alpha$, $\gamma$ and $\Gamma$ for the MIT and the NJL models, the radius $R^*$ is within the same order of magnitude in both cases (some hundreds of fm).  


In summary, we find that  fluctuations in the temperature and in the chemical potential of trapped neutrinos are not very important for deconfinement in neutron stars, and fluctuations in the energy density are the more efficient way to trigger the transition. Typically, over-densities  of $\delta \rho^H/\rho^H \sim 0.001-0.1$ above the bulk point are needed for the nucleation of drops with $R \sim 10-1000$ fm. However, the nucleation rates $\Gamma$ vary over several orders of magnitude.  Drops with radii between few fm to some hundreds of fm  have a huge nucleation rate while larger ones are strongly suppressed. 
In the context of neutron stars the main consequence is that if the bulk transition point is attained near the star center, quark matter drops with radii smaller than hundreds of fm will nucleate almost instantaneously.

\subsection{Finite size effects and the internal structure of compact stars}

Finite size effects are also determinant for compact stellar structure. A mixed phase of hadrons and quarks may arise only if the surface tension is smaller than a critical value of the order of tens of  MeV / fm$^2$  \cite{Voskresensky2003,Tatsumi2003,Maruyama2007,Endo2011}. Also,  surface tension affects decisively the properties of the most external layers of a strange star which may fragment into a charge-separated mixture, involving positively-charged strangelets immersed in a negatively charged sea of electrons, presumably forming a crystalline solid crust \cite{Jaikumar2006}. This would happen below a critical surface tension which is typically of the order of a few MeV/fm$^2$ \cite{Alford2006}.

In spite of its key role, the surface tension of quark matter is still poorly constrained. While early estimates pointed to values  below 5 MeV/fm$^2$ \cite{Berger1987}, larger values within  10 $-$ 50 MeV/fm$^2$ were used in other works about quark matter droplets in neutron stars \cite{Heiselberg1993,Iida1998}. More recently,  values around $\approx 30$ MeV/fm$^2$ have been adopted for studying the effect of quark matter nucleation on the evolution of protoneutron stars \cite{Bombaci2007,Bombaci2009}. However, much larger values have also been reported in the literature. Estimates given in  Ref. \cite{Voskresensky2003} give values in the range 50 $-$ 150 MeV/fm$^2$ and  values around  $\sim$ 300  MeV/fm$^2$ were suggested on the basis of dimensional analysis of the minimal interface between a color-flavor locked phase and nuclear matter \cite{Alford2001}.

In a recent work we evaluated the surface tension and the curvature energy of three-flavor quark matter  within the Nambu-Jona-Lasinio (NJL) model \cite{Lugones2013}. Finite size effects were addressed within the multiple reflection expansion formalism and  the effect of color superconductivity was taken into account.  We are interested in finite size color superconducting droplets  that may form  e.g. within the mixed phase of a cold hybrid star. In such a case, chemical equilibrium is maintained by weak interactions among quarks, and  therefore we have $\mu_{dc} = \mu_{sc} = \mu_{uc} + \mu_e$, for all colors $c$. 
We find that for $\mu$ below  $\sim 350$ MeV the solution that  minimizes the thermodynamic potential is the one corresponding to the chiral symmetry broken phase and  for $\mu$ above $\sim 350$ MeV it is the 2SC phase. For temperatures between 0$-$30 MeV and drop radii between 5$- \infty$ fm,  the surface tension is in the range of $\alpha \sim 145-165$ MeV/fm$^2$ and the curvature energy is in the range of $\gamma \sim 95-110$ MeV/fm.  The branches corresponding  to the chiral symmetry broken phase and the 2SC phase are separated by a discontinuity, but the change in the values of $\alpha$ and $\gamma$ is less than $\sim 10$ \% \cite{Lugones2013}. We also find that the temperature dependence of $\alpha$ and $\gamma$ is very weak in the temperature regime that is relevant for both protoneutron stars and cold neutron stars.
In the previous subsection we shown $\alpha$ and $\gamma$  for just deconfined quark matter \textit{out of chemical equilibrium} under weak interactions. Notice that, in spite of the very different thermodynamic conditions, the range of values of the surface tension $\alpha$ for quark matter in chemical equilibrium is similar, typically  $\sim 20$ \% larger. 

For such large values of $\alpha$ and $\gamma$,  the energy cost of forming quark drops within the mixed phase of hybrid stars would be very large.  According to \cite{Voskresensky2003}, beyond a limiting value of $\alpha \approx 65$ MeV/fm$^2$ the structure of the mixed phase becomes mechanically unstable  and local charge neutrality is recovered. Therefore, our results indicate that the hadron-quark interphase within a hybrid star should be a sharp discontinuity.

It is worth emphasizing that the nucleation of quark drops analyzed in the previous subsection  is possible even for large values of the surface tension. Since these drops are charge neutral due to flavor conservation \cite{Lugones2011},  they can be considerably larger than the Debye screening length $\lambda_D$ of the stellar plasma which is typically $5 - 10$ fm. For such large drops ($R \sim$ hundreds of  fm \cite{Lugones2011,doCarmo2013}), surface and curvature corrections tend to vanish. This is not the case for droplets of quark matter in the hypothetical mixed phase of a hybrid star. Since charge neutrality is a global condition in the mixed phase, quark drops are electrically charged and their size cannot exceed  $\sim \lambda_D \sim \mathrm{few \; fm}$. Therefore, finite size effects don't preclude the deconfinement nucleation but inhibits the formation of the mixed phase.

\section{Combustion processes: the growth of the initial drop}

\subsection{Processes inside the flame}

As described before, the formation of quark matter in a neutron star begins when  hadronic matter in the stellar core  attains a critical density and a small deconfined drop is nucleated (see Fig. \ref{fig_deconf_1}). The initial drop decays by means of weak interactions and as a result  a  quark drop in chemical equilibrium is formed, temperature is significantly increased, and neutrinos are produced. 

The energy released in such conversion can compress the adjacent hadronic matter, triggering the conversion  in the neighborhood of the drop. If the process is sufficiently exothermic, it may lead to the creation of a combustion front (flame) that travels outwards along the star and converts to quark matter the stellar core and even the whole star if quark matter is absolutely stable  \cite{Lugones2002,Keranen2005,Niebergal2010,Herzog2010,Fischer2011,Pagliara2013}.

\begin{figure}[tb]
\begin{center}
\resizebox{0.5 \textwidth}{!}{ \includegraphics[scale=0.5]{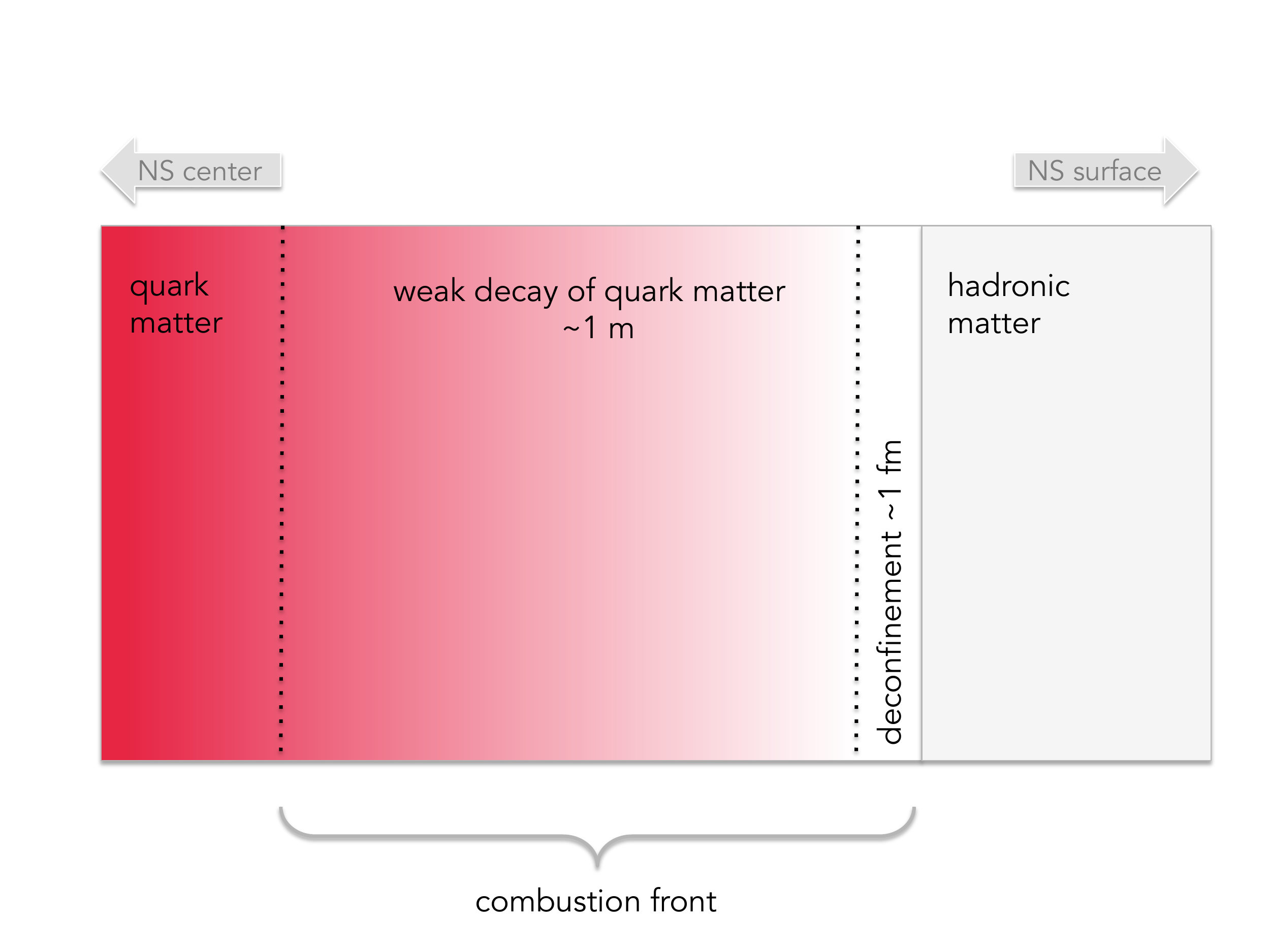} }
\end{center}
\caption{Structure of the flame that converts hadronic matter into quark matter in a compact star. In this figure the flame moves to the right. The front of the flame propagates with a  velocity of the order of $c/\sqrt{3}$, where $c$ is the speed of light  \cite{Lugones1994}. At the flame front there is a region of thickness $l_{strong} = \tau_{strong} \times c/\sqrt{3} = 10^{-23} \mathrm{s} \times  c/\sqrt{3} \sim 1 \mathrm{fm}$ where hadronic matter deconfines. Behind it, there is a region  of thickness $l_{weak} = \tau_{weak} \times c/\sqrt{3} = 10^{-8} \mathrm{s} \times  c/\sqrt{3} \sim 1 \mathrm{m}$ where quark matter reaches chemical equilibrium through weak interactions. } 
\label{fig_flame}
\end{figure}

According to this picture, the structure of the flame is as shown in Fig. \ref{fig_flame}. Deconfinement occurs within a thin layer of thickness $l_{strong} \sim 1 \mathrm{fm}$ behind which there is a thicker layer of about 1 m where the following processes drive quark matter into chemical equilibrium: 
\begin{eqnarray}
&& d  \rightarrow  u+e^{-}+\bar{\nu}_{e},\\
&& s  \rightarrow  u+e^{-}+\bar{\nu}_{e},\\
&& u+e^{-}  \leftrightarrow  d+\nu_{e},\\
&& u+e^{-}  \leftrightarrow  s+\nu_{e}. \\
&& u+d  \leftrightarrow  u+s,
\end{eqnarray}

The evolution of quark matter towards chemical equilibrium within the flame  can be described by 
\begin{eqnarray}
\frac{dY_u}{dt} = \frac{1}{n_b}   \sum_i   \Gamma^u_i , \\
\frac{dY_d}{dt} = \frac{1}{n_b}   \sum_i   \Gamma^d_i ,
\end{eqnarray}
where $\Gamma^f_i$ ($f = u, d$) are the rates of the processes creating and destroying the $f$ species.   Baryon number conservation and charge neutrality relate $s$ quark and electron abundances with $u$ and $d$ quark abundances:
$Y_{s}=3-Y_{u}-Y_{d}$, $Y_{e}=Y_{u}-1$.  For a transition to quark matter taking place in a protoneutron star, the lepton number $Y_{L}=Y_{e}+Y_{\nu}$ is constant because neutrinos are trapped. The evolution of the temperature can be obtained from the first law of thermodynamics \cite{Rosero2015,Anand1997,Dai1995}. 

The above equations can be solved using as initial conditions for the quark abundances the results of the previous section. The results  show that within the flame weak decays increase  significantly  the temperature (up to 60$-$70 MeV) and the strange quark abundance $Y_s$ in a timescale of $\sim 10^{-9}$ s \cite{Rosero2015}. The abundances of the other particles $Y_u$, $Y_d$, $Y_e$ decrease.   The nonleptonic process $u+d \rightarrow u+s$ is always dominant for protoneutron stars and cold neutron stars, but, in the case of  protoneutron star matter the process $u + e^{-} \leftrightarrow d + n_{\nu_e}$ becomes relevant as matter approaches chemical equilibrium.  The rates for all the other processes are orders of magnitude smaller. The electron capture reactions have a small contribution to the rate, but they are the most important processes for neutrino emission. The neutrino emissivity per baryon is very high during the first $\sim 10^{-10}$ s with a maximum value  in the range  $10^{10} -10^{12}$ MeV/s. After this initial phase, there is a steep decline to a value several  orders of magnitude smaller in a timescale of $\sim 10^{-8}$ s. 

Integrating in time the total neutrino emissivity we obtained  the total energy per baryon released by quark matter in the form of neutrinos \cite{Rosero2015}.  For cold neutron stars the energy release is $E_{\nu_{e}} = 30-60$ MeV per baryon and for protoneutron stars it is   $E_{\nu_{e}} = 10-55$ MeV.   Since a typical neutron star has $10^{58}$ baryons, the whole conversion of a compact object would release roughly  $\sim 10^{53}$ erg in the form of neutrinos due to the weak decays of quarks.
If the conversion to quark matter occurs in a protoneutron star,  these neutrinos can be absorbed by the matter just behind the shock wave that travels along the external layers of the progenitor star, and help to a successful core collapse supernova explosion. Notice also that the liberated energy is of the order of the energy of a gamma ray burst (GRB), indicating that models of GRBs involving the hadron to quark conversion in a neutron star deserve further study.

\subsection{Hydrodynamic evolution}

The study of the hydrodynamic evolution of the combustion front is an essential element not only for understanding the conversion timescale of the star, but also for assessing the potential observational signatures, e.g. in neutrinos, gravitational waves and gamma rays.  
Most works have concentrated on the hydrodynamics of the combustion under the assumption that quark matter is absolutely stable, i.e. that it has an energy per baryon at zero temperature and vanishing pressure that is lower than the mass of the neutron (the so called Bodmer-Witten- Terazawa hypothesis \cite{Witten1984}).  If such hypothesis is true, strange quark matter would be the true ground state of strongly interacting matter and therefore all hadronic neutron stars would be in a metastable state.  
Early papers \cite{Olinto1991,Heiselberg1991} have estimated the velocity of the laminar deflagration by considering the diffusion of $s$-quarks as the main agent for the progress of the conversion, finding 
 that the laminar velocity of the front is relatively slow ($v_{lam} < 10^4$ cm/s). This velocity is a direct consequence of both the timescale for weak decays that create $s$-quarks  and the physics of the diffusion process.  In more recent studies assuming strange quark diffusion \cite{Niebergal2010},  the combustion of pure neutron matter to strange quark matter is numerically investigated taking into account the binding energy release and neutrino emission across the burning front.  In such calculations the typical speed of the burning process is between 0.002 $c$ and 0.04 $c$.  However, it is well known that deflagration fronts are subject to several instabilities (e.g. Landau-Darrieus and Rayleigh-Taylor), that lead to a wrinkling of the flame and an increased combustion area that may cause a strong flame acceleration depending on the role of non-linear stabilizing effects.  Whether the conversion process remains  as a deflagration, either laminar or turbulent, or jumps to the detonation regime, thus driving an explosive transient  has been debated in the literature  for many years \cite{Benvenuto1989,Lugones1994,Lugones2002,Tokareva2004,Keranen2005,Tokareva2006,Drago2007,Niebergal2010,Herzog2010,Fischer2011,Pagliara2013}.  A definitive answer to these questions  would need  full numerical simulations involving general relativistic hydrodynamics and a detailed description of the equations of state, neutrino interactions and neutrino transport.  Some work has been done in this direction, but the complexity of the problem demands strong approximations. In Refs. \cite{Herzog2010,Pagliara2013}  the authors present three-dimensional non-relativistic numerical simulations of turbulent combustion and  in all cases they observe growing Rayleigh-Taylor instabilities of the conversion front. The resulting turbulent motion strongly enhances the conversion velocity but the burning speed does not reach sonic or even supersonic velocities. In contrast, preliminary one-dimensional relativistic hydrodynamic simulations point to supersonic conversion fronts \cite{Albarracin2015}.

\section{Structure of quark stars:  hybrid and strange configurations}

The recent determination of the mass of the pulsars PSR J1614-2230 with $M = (1.97 \pm 0.04) M_{\odot}$ \cite{Demorest2010} and  PSR J0348-0432 with $M = (2.01 \pm 0.04) M_{\odot}$ \cite{Antoniadis2013}  renewed the discussions about the possibility of exotic matter being present at  the core of neutron stars.  Such interest is increased by the possible existence of even more massive neutron stars such as the black-widow pulsar PSR J1311-3430 \cite{Romani2012} and  PSR J1816+4510 \cite{Kaplan2013}.

One  of the problems that reemerged is the softening of the equation of state of dense baryonic matter due to the presence of hyperons at high density, which lowers the maximum mass of hadronic stars. This led to the exploration of  additional repulsion between hyperons due to a vector meson coupled to hyperons only, as well as the possible appearance of  a transition to a very stiff quark phase before the hyperon threshold (see \cite{Buballa2014} and references therein). Concerning quark matter, it is known that schematic MIT bag models of strange stars made of absolutely stable quark matter satisfy comfortably the new constraint if color-superconductivity is taken into account \cite{Lugones2003,Horvath2004}. However, it is not straightforward to construct models of hybrid stars with more than two solar masses.  If the quark phase is described within the MIT bag model,  hybrid stars reach the two solar mass limit only for a significantly limited range of parameters of the models
that include corrections motivated from perturbative one-gluon exchange or from color superconductivity. Moreover, the appearance of quark matter in massive stars crucially depends on the stiffness of the nuclear matter equation of state \cite{Weissenborn2011}.

The situation is similar if the quark phase is described by the NJL model. In fact, in a recent  work  we performed a systematic study of hybrid star configurations using a relativistic mean-field hadronic equation of state \cite{Walecka1974,Glendenning1991}  and the NJL model for three-flavor quark matter. For the hadronic phase we used the stiff GM1 and TM1 parametrizations, as well as the very stiff NL3 model \cite{Lenzi2012}. In the  NJL Lagrangian we included scalar, vector and 't Hooft interactions, with the vector coupling constant $g_v$ treated as a free parameter.  Within this approach, the hadronic and the quark-gluon degrees of freedom are derived from different Lagrangians and the quark-hadron phase transition is associated with the point where both models have the same free energy. Thus, by construction, chiral symmetry restoration in the quark model occurs at a chemical potential $\mu$ that is in general different to the $\mu$ of deconfinement. Additionally, the thermodynamic potential within the NJL model is defined but for a 'bag constant'  $\Omega_0$, analogous to the MIT bag constant, which is usually fixed by requiring that the corrected pressure  is vanishing at vanishing chemical potential \cite{Schertler1999,Buballa1999,Blaschke2005}. Although this has been the most commonly employed procedure, it is completely arbitrary an in fact different approaches have been adopted in the literature \cite{Pagliara2008,Lenzi2012,Bonanno2012}. For example, in \cite{Pagliara2008} it is assumed that deconfinement occurs at the same chemical potential as the chiral phase transition, i.e. the value of $\Omega_0$ is fixed to force the pressure of quark matter to be equal to the pressure of the hadronic matter at the critical chemical potential for which chiral symmetry is restored. The bag value obtained with this assumption is the lowest possible value for the bag constant in the NJL model because it allows to use the NJL equation of state just starting from the chemical potential of the chiral phase transition. Using this procedure, \cite{Pagliara2008} computed the equation of state of quark matter within the NJL model by including effects from the chiral condensates, the diquark coupling pattern, and a repulsion vector term. They find that hybrid stars containing a CFL core are stable but the maximum mass is $\sim 1.8 M_{\odot}$, i.e. incompatible with the recent observed pulsars.

More recently, we have explored the effect of different calibrations of the bag constant. Instead of $\Omega_0$, we used a value $\Omega_0 + \delta \Omega_0$, where $\delta \Omega_0$ is a free parameter. Notice that the parametrization of the model can be kept as in the standard case because a change in the value of $\Omega_0$ has no influence on the fittings of the vacuum values for the pion mass, the pion decay constant, the kaon mass, the kaon decay constant, and the quark condensates. The effect on the equation of state can be understood as follows: first, the chemical potential at which the chiral transition occurs doesn't depend on  $\Omega_0$  because  it is determined from the solution of the gap equations for the constituent masses. On the other hand, the chemical potential  for the deconfinement transition depends on  $\Omega_0$ because it is determined by matching the pressures of the hadronic and quark phases.  Thus, tuning $\delta \Omega_0$ is an easy way to control the splitting between both transitions. Clearly, $\delta \Omega_0$ has a minimum value because the phase  transition cannot be shifted to a pressure regime where the NJL model describes the vacuum. That is, we fix a minimum limit to $\delta \Omega_0$ for which the phase transition occurs at the chiral symmetry restoration point. On the other hand, there is no maximum value in principle for  $\delta \Omega_0$, since the phase transition can be shifted to arbitrarily large pressures.

The effect on the structure of spherically symmetric and static stars  is the following. As we increase the value of $\delta \Omega_0$, the deconfinement pressure gets larger, leading to configuration with smaller quark matter cores.
At the same time, we obtain larger maximum masses because the hadronic equation of state is stiffer than the quark one. However, since there is a larger density jump between the two phases, the configurations tends to be less stable; i.e. hybrid stars  are stable within a smaller range of central densities. For large enough $\delta \Omega_0$, stable hybrid configurations are not possible at all \cite{Lenzi2012}. The effect of increasing the coupling constant $g_v$ is very similar.  When hyperons are included in the hadronic equation of state, the main effect  is that they preclude the deconfinement transition in almost all the region of the parameter space that allows large maximum masses.
In summary,  hybrid configurations with maximum masses above $\sim 2 M_{\odot}$ are possible for a significant region of the parameter space of $g_v$ and  $\delta \Omega_0$ provided a stiff enough hadronic equation of state without hyperons is used. However,  the observation of compact star masses a few percent larger than the mass of PSR J1614-2230 will be hard to explain within hybrid star models using the GM1 and TM1 equations of state and will require a very stiff hadronic model such as NL3 with nucleons only.

\section{Signatures of quark matter from compact star pulsation modes}

A new window for the observation of compact stars will be opened soon thanks to the second generation of interferometric gravitational wave detectors (Advanced LIGO, Advanced Virgo and KAGRA \cite{Riles2013}). When the design sensitivity is reached, which is expected to happen around the year 2021,  such instruments will be ten times more sensitive than the first generation and will reach about $10^5$ galaxies allowing a NS-NS merging detection rate of $\sim 1$ per month. 
Such sensitivity will probably allow the detection of pulsation modes of compact stars excited in binary mergers or pulsations of newly born pulsating compact objects associated with the violent dynamics of core collapse supernovae. 

Lot of work has been done in the last four decades in order to describe the non-radial oscillatory properties of neutron stars \cite{Lindblom1985,Yoshida1997,Andersson1998,Yip1999,Sotani2001,Miniutti2003,Benhar2004,Benhar2007,Sotani2011,Flores2014};	
however, these studies employed equations of state that in most cases render maximum stellar masses below $2 M_{\odot}$ i.e. inconsistent with the recently observed high-mass pulsars. Therefore, it is worth re-examining the oscillation spectra because the change in the allowed equation of state may bring new ways to distinguish hadronic, hybrid and strange stars. 

In  a recent paper \cite{Flores2014}  we have investigated non-radial fluid oscillations of hadronic, hybrid and strange quark stars with maximum masses above   $2 M_{\odot}$.  For the hadronic equation of state we employed two different parametrizations of a relativistic mean-field model with nucleons and electrons. For quark matter we have included the effect of strong interactions and color superconductivity within the MIT bag model. The equations of non-radial oscillations were integrated within the Cowling approximation in order to determine the frequency of the fundamental ($f$) mode, the first pressure ($p_1$) mode, and the discontinuity gravitational ($g$) mode of hybrid stars.

We find that the fundamental mode is sensitive to the internal composition, but due to the uncertainties in the equations of state, there is an overlapping at $\sim 2$ kHz of the curves corresponding to hadronic, hybrid and strange quark stars for stellar masses larger that $ \sim 1 ~ M_{\odot}$.  As a consequence it would be difficult to distinguish hybrid and hadronic stars through the $f$-mode frequency, even if the mass or the surface $z$ of the object is determined concomitantly with $f_f$. However, there are features that in some cases may allow a differentiation between strange stars and hadronic/hybrid stars. For example, strange stars cannot emit gravitational waves with frequency below $\sim 1.7$ kHz for any value of the mass. Also, sources with a mass in the range $1-1.5 M_{\odot}$ emitting a signal in the range $2-3$ kHz would be strange stars. 

The frequency of the $p_1$ mode is much more affected by the internal composition of the star.  For hadronic and hybrid stars, we find that $f_{p1}$ is in the range $4-7$ kHz for objects with masses in the range $1-2 \, M_{\odot}$, but for strange quark stars it can be significantly larger than $\sim 7$ kHz. Thus, a compact object emitting a signal above $\sim 7$ kHz could be identified as a strange star even if its  mass or gravitational redshift are unknown. 

Discontinuity $g$-modes are only present in hybrid stars with sharp discontinuities and fall in the range $0.4-1$ kHz. Thus, they are clearly distinguishable from the $f$ mode, and of low frequency $g$-modes associated with 
chemical inhomogeneities in the outer layers or thermal profiles. Our results are summarized in Table \ref{table2} and show that based on the simultaneous analysis of the frequency of the $f$, $p_1$ and $g$-modes it would be possible to discriminate between hadronic, hybrid and strange quark stars.

%
\begin{table}
\centering
\begin{tabular}{l|ccc}
\hline 
            &       $f_f$        & $f_{p1}$   & $f_{g}$   \\
\hline 
strange stars  &       $\sim 2$ kHz &  $>  7$ kHz    &   not present   \\
hybrid stars   &       $\sim 2$ kHz &  $\sim 4-7$ kHz    &  $\sim  0.4-1$ kHz \\
hadronic stars &       $\sim 2$ kHz &  $\sim 4-7$ kHz    &   not present   \\
\hline
\end{tabular}
\caption{Discrimination between hadronic, hybrid and strange quark stars based on the observation of the frequency of the $f$, $p_1$ and $g$ modes.} 
\label{table2}
\end{table}

\section{Summary and conclusions}

We reviewed some recent studies  about  the nucleation of quark matter droplets under neutron star and protoneutron star conditions.   The direct nucleation of a quark matter drop in equilibrium under weak interactions is
strongly suppressed because it involves the simultaneous conversion of many non-strange quarks into strange quarks.
Thus, an activation state with larger free energy than the final state must be nucleated first.  Since such state is reached through strong interactions,  flavor is conserved and the initial quark matter drop is out of chemical equilibrium with respect to weak interactions.  Under these assumptions, the surface tension and curvature energy of quark matter are calculated self-consistently within the multiple reflection expansion formalism for different quark matter equations of state (MIT bag model and NJL model). We also determine the transition density for different drop sizes as a function of temperature and show that density fluctuations are much more relevant than temperature fluctuations for initiating the conversion.  A quite general result is that the  nucleation rate for density fluctuations is negligible for droplets larger than some hundreds of fm  and is huge for smaller droplets. In spite of some quantitative differences, this result has been shown to be valid  for calculations using the Nambu-Jona-Lasinio model  and the MIT bag model. Thus, we conclude that small drops of quark matter must be almost instantaneously nucleated at the core of a compact star when the bulk transition density is attained due to accretion, spin down or any other astrophysical mechanism. 

Once such drops are formed, they may grow if the conversion is sufficiently exothermic because the released energy  can compress the adjacent hadronic matter to a density above the critical one. After some transient, a  burning front may be created with a thickness of $\sim 1 \mathrm{m}$ where the temperature rises up to 60$-$70 MeV and the integrated neutrino emissivity  is in the range $E_{\nu_{e}} = 10-60$ MeV per baryon. An important issue is whether the combustion proceeds in a fast or a slow hydrodynamic  mode, because this is closely related to the possible observational outcome of the conversion.  A definitive answer  would need  full numerical simulations involving general relativistic hydrodynamics and  detailed microphysics which is still missing in the literature, as was discussed in Section 3.  

We also analyzed the structure of hybrid stars within the Nambu-Jona-Lasinio model.  It is shown that  for  three-flavor quark matter in chemical equilibrium the surface tension is in the range of $\alpha \sim 145-165$ MeV/fm$^2$ and the curvature energy is in the range of $\gamma \sim 95-110$ MeV/fm. For such large values of $\alpha$ mixed phases are mechanically unstable and the hadron-quark interphase in a hybrid star should be a sharp discontinuity. The structure of hybrid stars is explored for a NJL model  with scalar, vector and 't Hooft interactions,  paying particular attention to the role of vector interactions and  to a non-standard  choice of  the 'bag constant'  which  is usually fixed by requiring that the corrected pressure is vanishing at vanishing chemical potential. Taking the vector coupling constant and the 'bag constant' as free parameters 
we find that hybrid configurations with maximum masses above $\sim 2 M_{\odot}$ are possible for a significant region of the parameter space  provided a stiff enough hadronic equation of state without hyperons is used. 

Finally, we have briefly described the non-radial pulsation properties of hadronic, hybrid and  strange quark stars with maximum masses above $2 M_{\odot}$ using several equations of state. Such analysis is relevant because future generations of gravitational wave detectors will probably allow the identification of normal oscillation modes excited in catastrophic astrophysical events.  Important information about the existence of quark matter phases can be obtained from the $p_1$ and the $g$ mode. While a  value of the $p_1$ mode above $\sim 7$ kHz would point that the emitting object is a strange quark star, the observation of a signal in the $0.4-1$ kHz range would be indicative of a $g$ mode associated  with  the sharp quark-hadron discontinuity in a hybrid star.

\section*{Acknowledgements}   I am grateful to C\'esar V\'asquez Flores, Gabriela Grunfeld, Gustavo Colvero,  Jhon Anderson Rosero, Marcos Albarracin Manrique and Taiza do Carmo for fruitful discussions on the topics of the present work.  I acknowledge the Brazilian agencies FAPESP and CNPq  for financial support.


\begin{thebibliography}{}




\bibitem{Demorest2010} P. B. Demorest  et al., Nature 467, 1081 (2010).

\bibitem{Antoniadis2013} J. Antoniadis   et al., Science 340, 1233232 (2013). 

\bibitem{Romani2012} R. W. Romani  et al.,  Astrophys. J. 760, L36 (2012).

\bibitem{Kaplan2013} D. L. Kaplan  et al.,  Astrophys. J. 765, 158 (2013).



\bibitem{Balian1970} R. Balian and C. Bloch,  Annals of Physics 60, 401  (1970).

\bibitem{Madsen1994}  J. Madsen, Phys. Rev. D {50}, 3328 (1994).

\bibitem{Kiriyama2003} O. Kiriyama and A. Hosaka, Phys. Rev.  D {67}, 085010 (2003).

\bibitem{Kiriyama2005} O. Kiriyama, Phys. Rev.  D {72}, 054009 (2005).

\bibitem{Lugones2011}	G. Lugones and A. G. Grunfeld, Phys. Rev. D 84, 85003 (2011).

\bibitem{Lugones2013}	G. Lugones, A. G. Grunfeld, and M. Al Ajmi, Phys. Rev. C 88, 045803 (2013).

\bibitem{Lugones2010}	G. Lugones, T. A. S. do Carmo, A. G. Grunfeld, and N. N. Scoccola, Phys. Rev.  D 81, 85012 (2010).

\bibitem{Lugones2009}	G. Lugones, A. G. Grunfeld, N. N. Scoccola, and C. Villavicencio, Phys Rev D 80, (2009).

\bibitem{doCarmo2013b}	T. A. S. do Carmo and G. Lugones, Physica A 392, 6536 (2013).

\bibitem{Lugones2005} G. Lugones and I. Bombaci, Phys. Rev. D { 72}, 065021 (2005).

\bibitem{Walecka1974} J.D. Walecka, Ann. Phys. { 83}, 491 (1974);  B.D. Serot and J.D. Walecka, Adv. Nucl. Phys. { 16}, 1 (1986).

\bibitem{Glendenning1991} N. K. Glendenning and S.A. Moszkowski, Phys. Rev. Lett. { 67}, 2414 (1991).

\bibitem{Logoteta2009} I. Bombaci, D. Logoteta, P. K. Panda, C. Providencia, I. Vidana, Phys.  Lett. B {680}, 448 (2009).

\bibitem{prefactor} J. S. Langer, Phys. Rev. Lett. 21, 973  (1968);  J. S. Langer, Ann. Phys. (N.Y.) 54, 258 (1969);  J. S. Langer, L. A. Turski, Phys. Rev. A 8, 3230  (1973);  L. A. Turski, J.S. Langer, Phys. Rev. A 22, 2189 (1980); L. Csernai, J. I. Kapusta, Phys. Rev. D 46, 1379 (1992);  R. Venugopalan, A. P. Vischer, Phys. Rev. E 49, 5849 (1994).

\bibitem{doCarmo2013} T. A. S. do Carmo, G. Lugones, and A. G. Grunfeld, J. Phys.  G: Nucl. Part. Phys. 40, 035201 (2013).


\bibitem{Voskresensky2003}	D. N. Voskresensky, M. Yasuhira, and T. Tatsumi, Nucl. Phys. A 723, 291 (2003).

\bibitem{Tatsumi2003}	T. Tatsumi, M. Yasuhira, and D. N. Voskresensky, Nucl. Phys. A 718, 359 (2003).

\bibitem{Maruyama2007}		T. Maruyama, S. Chiba, H.-J. Schulze, and T. Tatsumi, Phys. Rev. D 76, 123015 (2007).

\bibitem{Endo2011}		T. Endo, Phys. Rev. C 83, 068801 (2011).

\bibitem{Jaikumar2006} P. Jaikumar, S. Reddy, and A. W. Steiner, Phys. Rev. Lett. 96, 041101 (2006).

\bibitem{Alford2006}	M. Alford, K. Rajagopal, S. Reddy, and A. Steiner, Phys. Rev. D 73, 114016 (2006).

\bibitem{Berger1987} M. S. Berger and R. L. Jaffe, Phys. Rev. C 35, 213 (1987); Phys. Rev. C 44, R566 (1991)

\bibitem{Heiselberg1993}	H. Heiselberg, C. J. Pethick, and E. F. Staubo, Phys. Rev. Lett. 70, 1355 (1993).

\bibitem{Iida1998} K. Iida and K. Sato, Phys. Rev. C { 58}, 2538 (1998).

\bibitem{Bombaci2007}   I. Bombaci, G. Lugones, I. Vida\~na, Astron.  Astrophys. { 462}, 1017 (2007).

\bibitem{Bombaci2009} I. Bombaci, D. Logoteta, P.K. Panda, C. Providencia, I. Vidana,  Phys. Lett. B {680}, 448 (2009).

\bibitem{Alford2001} M. G. Alford, K. Rajagopal, S. Reddy, and F. Wilczek, Phys. Rev. D 64, 074017 (2001).

\bibitem{doCarmo2013}	T. A. S. do Carmo, G. Lugones, and A. G. Grunfeld, J. Phys. G: Nucl. Part. Phys. 40, 035201 (2013).




\bibitem{Rosero2015} J. A. Rosero and G. Lugones, to appear in Nucl. Phys. Proc. Suppl.  (2015).

\bibitem{Anand1997} J. D. Anand, A. Goyal, V. K. Gupta and S. Singh, {Astrophysical J.} { 481}, 954 (1997)

\bibitem{Dai1995} Z. Dai, Q. Peng, and T. Lu,  {Astrophys. J.} {440}, 815, (1995).

\bibitem{Witten1984} 	E. Witten, Phys. Rev.  D 30, 272 (1984).

\bibitem{Olinto1991} A. V. Olinto, Nucl. Phys. Proc. Suppl. 24B, 103 (1991).

\bibitem{Heiselberg1991} H. Heiselberg, G. Baym, and C. J. Pethick, Nucl. Phys. Proc. Suppl. 24B, 144 (1991).

\bibitem{Benvenuto1989} O. G. Benvenuto and J. E. Horvath,  Phys. Rev. Lett.  63,  716 (1989).

\bibitem{Lugones1994}  	G. Lugones, O. G. Benvenuto, and H. Vucetich,  Phys. Rev. {D50}, 6100 (1994).

\bibitem{Lugones2002} G. Lugones, C. R. Ghezzi, E. M. de Gouveia Dal Pino, and J. E. Horvath,  Astrophys. J. 581, L101 (2002).

\bibitem{Tokareva2004} I. Tokareva, A. Nusser, V. Gurovich, and V. Folomeev, {Int. J. Mod. Phys.} {D14}, 33 (2005).

\bibitem{Keranen2005} P. Keranen, R. Ouyed, and P. Jaikumar,  {Astrophys. J.} {618}, 485  (2005).

\bibitem{Tokareva2006} I. Tokareva and A. Nusser, {Phys. Lett.} {B639}, 232 (2006).

\bibitem{Drago2007} A. Drago, A. Lavagno, and I. Parenti, Astrophys. J.  659, 1519 (2007).

\bibitem{Niebergal2010} B. Niebergal, R. Ouyed, and P. Jaikumar,  Phys. Rev.  {C 82}, 062801 (2010).

\bibitem{Herzog2010}	M. Herzog and F. K. R\"opke,  Phys. Rev.   {D 84}, 083002 (2011).

\bibitem{Fischer2011} T. Fischer, et al.  {Astrophys. J. Supp.} {194}, 39 (2011).
 
\bibitem{Pagliara2013}	G. Pagliara, M. Herzog, and F. K. R\"opke,  Phys. Rev.   {D 87}, 103007 (2013).

\bibitem{Albarracin2015}  M. A. Albarracin Manrique and G. Lugones, Braz. J. Phys. 45, 457 (2015).




\bibitem{Buballa2014} 	M. Buballa, V. Dexheimer, A. Drago, E. Fraga, P. Haensel, I. Mishustin, G. Pagliara, J. Schaffner-Bielich, S. Schramm, A. Sedrakian, and F. Weber, J. Phys. G: Nucl. Part. Phys. 41 (2014) 123001.

\bibitem{Lugones2003} G. Lugones and J. E. Horvath, Astron. Astrophys. 403, 173 (2003).

\bibitem{Horvath2004}  J. E. Horvath and G. Lugones, Astron. Astrophys. 422, L1 (2004).

\bibitem{Weissenborn2011} S. Weissenborn, I. Sagert, G. Pagliara, M. Hempel, and J. Schaffner-Bielich, Astrophys. J. Lett. 740, L14 (2011).

\bibitem{Lenzi2012}	C. H. Lenzi and G. Lugones, Astrophys. J. { 759}, 57 (2012).

\bibitem{Schertler1999} K. Schertler, S. Leupold, and J. Schaffner-Bielich, Phys. Rev. C 60, 025801 (1999).

\bibitem{Buballa1999} M. Buballa and M. Oertel, Phys. Lett. B 457, 261 (1999).

\bibitem{Blaschke2005} D. Blaschke, S. Fredriksson, H. Grigorian, A.M. Oztas, and F. Sandin, Phys. Rev. D 72, 065020 (2005).

\bibitem{Pagliara2008} G. Pagliara and J. Schaffner-Bielich, {Phys. Rev.} D 77, 063004 (2008).

\bibitem{Bonanno2012}  	L. Bonanno and A. Sedrakian, Astron. Astrophys. 539, A 16 (2012).




\bibitem{Riles2013} K. Riles, Prog. Part. Nucl. Phys.  68, 1 (2013).

\bibitem{Lindblom1985} S. Detweiler and L. Lindblom, Astrophys. J. 292, 12 (1985).

\bibitem{Yoshida1997} 	S. Yoshida and Y. Kojima, Mon. Not. R. Astron. Soc. 289, 117 (1997).

\bibitem{Andersson1998} N. Andersson and K. D. Kokkotas, Mon. Not. R. Astron. Soc.  299, 1059 (1998).

\bibitem{Yip1999} C. W. Yip, M.-C. Chu, and P. T. Leung,  Astrophys.  J.  513, 849 (1999).

\bibitem{Sotani2001} H. Sotani, K. Tominaga, and K.-I. Maeda, Phys. Rev. D 65, 024010 (2001).

\bibitem{Miniutti2003}  G. Miniutti, J. A. Pons, E. Berti, L. Gualtieri, and V. Ferrari, Mon. Not. R. Astron. Soc. 338, 389 (2003).

\bibitem{Benhar2004}  O. Benhar, V. Ferrari, and L. Gualtieri, Phys. Rev. D 70, 124015 (2004).

\bibitem{Benhar2007}  O. Benhar, V. Ferrari, L. Gualtieri, and S. Marassi, Gen. Relat. Gravit. 39, 1323 (2007).

\bibitem{Sotani2011}	H. Sotani, N. Yasutake, T. Maruyama, and T. Tatsumi, Phys. Rev. D 24014 (2011).

\bibitem{Flores2014}	C. V. Flores and G. Lugones, Class. Quantum Grav. 31, 15502 (2014).


\end{thebibliography}
\end{document}